\documentclass{article}
\usepackage{graphicx}
\date{}
\begin{document}
\title{ 
PERSPECTIVES FOR MUON COLLIDERS AND NEUTRINO FACTORIES}
\author{M. Bonesini        \\
{\em  Sezione INFN Milano Bicocca, Dipartimento di Fisica G. Occhialini,}\\ 
{\em Universit\`a di Milano Bicocca, Milano, Italy.}
}
\maketitle
\baselineskip=11.6pt
\begin{abstract}
High brilliance muon beams are needed for  future facilities such
as a Neutrino Factory, an Higgs-factory or a multi-TeV Muon Collider.
The $R \& D$  path involves many aspects, of which 
cooling of the incoming muon beams is essential.
%% The main issues in this path will be outlined.
\end{abstract}
\baselineskip=14pt
\vskip -1cm

\section{Introduction}
\begin{figure}[htbp] % figures (and tables) should go top or bottom of
                    % the page where they are first cited or in
                    % subsequent pages
\begin{center}
\vskip -0.2cm
{\includegraphics[scale=0.50]{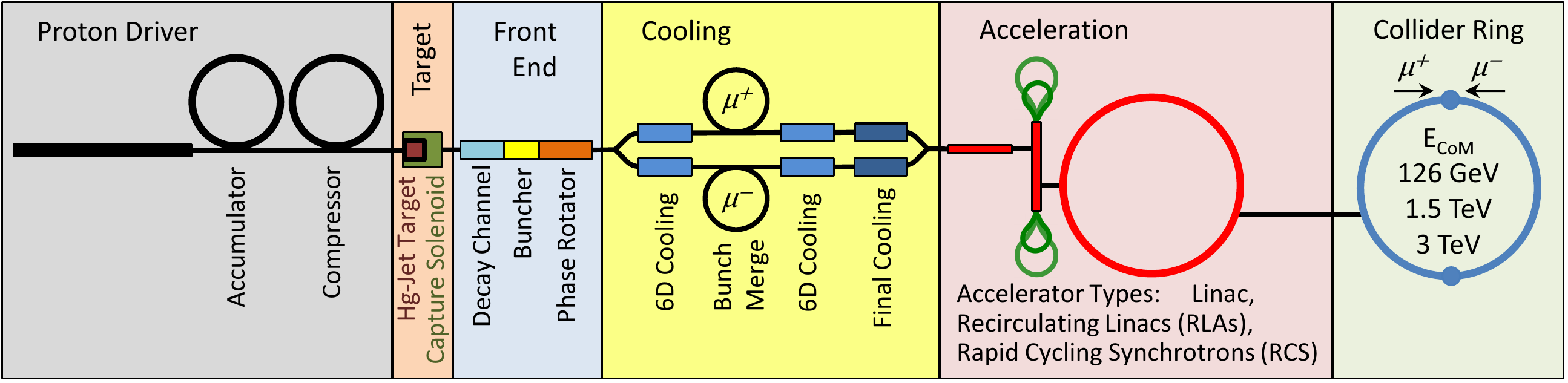}}
\vskip 0.5cm
{\includegraphics[scale=0.60]{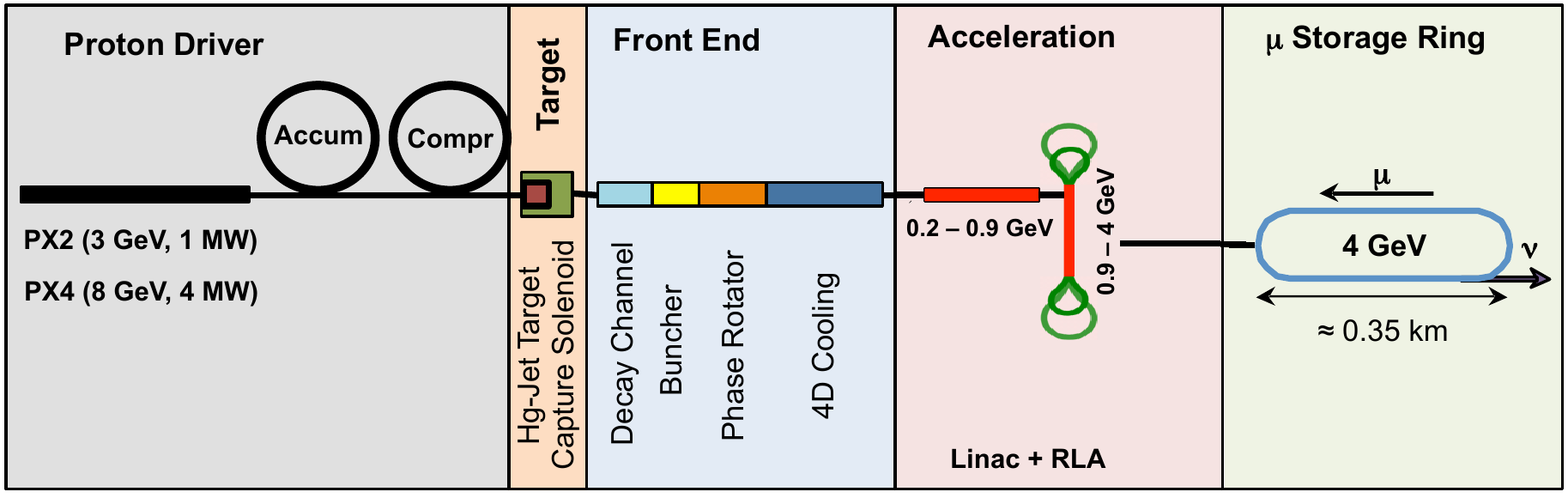}}
\end{center}
\caption{Schematic layout of a MC (top) and a NF (bottom).}
\label{fig:MC-NF}
\end{figure}

Since the 1960's,   
 Muon Colliders (MC) \cite{Tikhonin} 
and  Neutrino Factories (NF) \cite{Kosharev}, based on high
brilliance muon beams, have been proposed.
Their design has been optimized
 in references
\cite{Geer1}, \cite{US2a}, \cite{Choubey},\cite{Apollonio}  
and \cite{Bogomilov}.  
While a  MC addresses the high-energy frontier:
looking at precise Higgs physics \cite{Cline} and beyond, a
 NF  will provide the ultimate tool for neutrino oscillation studies,
looking at CP-violation.
The current design of a NF or a MC
 front-end is similar, up to the beginning of 
the cooling section , as can be seen from the 
layouts reported in figure \ref{fig:MC-NF}. 
%%In this way, a NF may be considered as a first step in a 
%%MC accelerator complex.

MC's may be developed with  c.m.s energy up to many TeV and, due to
the large $\mu$ mass as compared to the electron one, may easily 
fit in the footprint of existing HEP laboratories~\footnote{A $\sqrt{s}=3$ 
TeV Muon 
Collider ($\mu^+\mu^-$ Higgs Factory) has a ring circumference 
of $\sim 6.3 (\sim 0.3)$ km, to be compared
to the $\sim 26$  km  of the LHC tunnel.}

s-channel scalar Higgs production is greatly enhanced 
in a $\mu^+ \mu^-$
collider (as respect to $e^+ e^-$) as the coupling
is proportional to the lepton mass. Precision
measurements in the Higgs sector are thus feasible:  at $m_{H^0} \sim 126 $ 
GeV/$c^2$
only a $\mu^+ \mu^- $ collider may directly measure the $H^0$ lineshape.
With an integrated luminosity of 0.5 fb$^{-1}$,  
the $H^0$ mass may be determined, in the Standard Model case, 
 with a precision of 0.1 MeV$/c^2$ and
its width $\Gamma_{H^0}( \sim 4$ MeV$/c^2$)  with a precision of 0.5 MeV$/c^2$.

Through the processes $\mu^- \mapsto e^- \nu_{\mu} \overline{\nu_e}$ and
$\mu^+ \mapsto e^+ \overline{\nu_{\mu}} \nu_e$, neutrino beams 
with a flux known at better than $1 \%$ and well-known 
composition ($50 \% \nu_{\mu}$ or $\overline{\nu_{\mu}}$, $50 \% 
\overline{\nu_e}$
or $ \nu_e$) may be produced in a NF \cite{Bonesini1}. 
The ``golden channel'' linked to $\nu_e \mapsto \nu_{\mu} $ (or 
$\overline{\nu_e} \mapsto \overline{\nu_\mu}$) oscillations, manifests itself
by wrong sign muons, as respect to initial beam charge, suggesting  the use
of large magnetized far detectors.  
After the experimental discovery of a large $\theta_{13}$ value,
$\sim 5 \%$,
%% as reported in global fit of neutrino data of reference 
%%\cite{Gonzalez})
the design of the NF has been revised to improve 
precision in the study of sub-leading effects in neutrino oscillations
and provide better capabilities for the measurement of the phase $\delta$,
if leptonic CP-violation occurs \cite{Bogomilov}.
\section{R$\&$D towards a muon collider and a neutrino factory}
Many R$\&$D issues are relevant for the development of a NF or a MC, such as
the availability of a suitable proton driver or a high-power target, but
the most critical one is still the muon cooling.  
Muons are produced as tertiary particles in the process chain 
$p A \mapsto \pi X, \pi \mapsto \mu \nu$ and thus occupy a large 
longitudinal and transverse phase space. 
Conventional accelerator technologies  require
input beams with small phase space. 
To alleviate this problem one may use either new large
aperture accelerators, such as ``fixed field alternating-gradient'' (FFAG)
machines \cite{Rees} or try to reduce (``cool'') the 
incoming muon beam phase space.
While for a NF the required cooling factor is
small: around 2.4 for the 75 m cooling section
in the IDS-NF  design  \cite{Choubey}, \cite{Bogomilov},
for a MC
a longitudinal emittance reduction $\sim 14$  and a transverse 
emittance reduction $\sim 400$ in both transverse coordinates
 are needed, 
requiring a total cooling factor $\sim 2 \times 10^6$. 
%%This shows how
%%the engineering of the cooling section is critical  on the way
%%to a MC. 

%
\subsection{Ionization cooling and the MICE experiment at RAL}
Conventional beam cooling methods do not work on the short 
timescale of the muon lifetime ($\tau \sim 2.2 \mu$s). 
The only effective way is the 
so-called ``ionization cooling'' that is accomplished by passing
muons through a low-Z absorber, where they loose energy by 
ionization 
and the longitudinal component of momentum is then
replenished by RF cavities \cite{srinsky}. 

The initial goal of the MICE experiment~\cite{mice}  to study
a fully engineered cooling cell  of the 
proposed US Study 2~\cite{US2a}, has been
downsized in 2014 to a demonstration of ionization cooling with a simplified
lattice based on the
available RF cavities and absorber-focus coils (see the top panel 
of figure 2). 
\begin{figure}[h]
\begin{center}
{\includegraphics[scale=0.32]{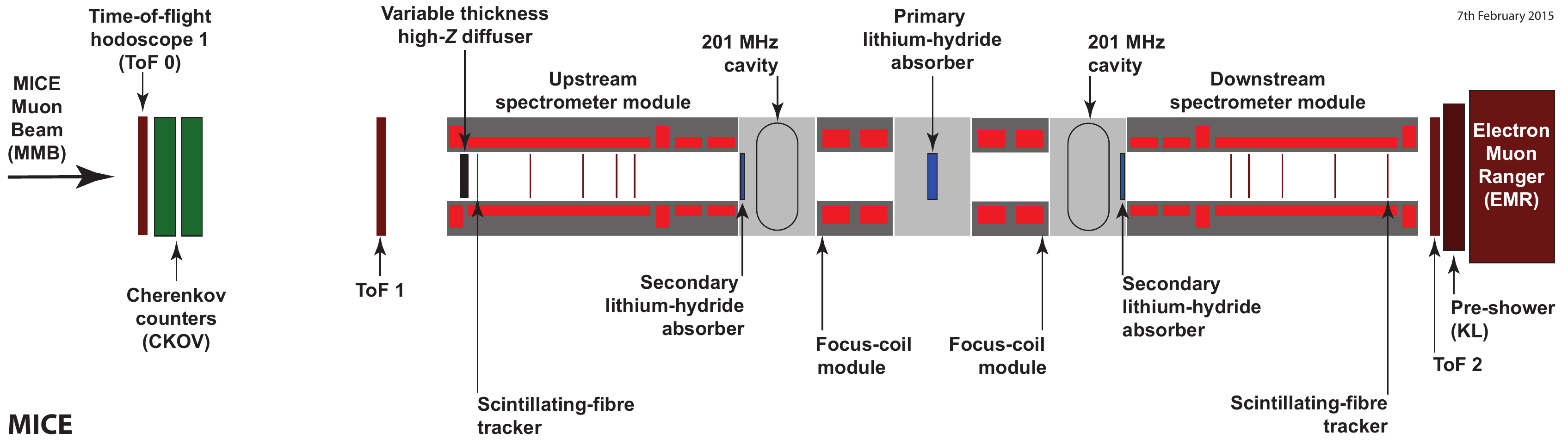}}
\vskip -0.2cm
{\includegraphics[scale=0.34]{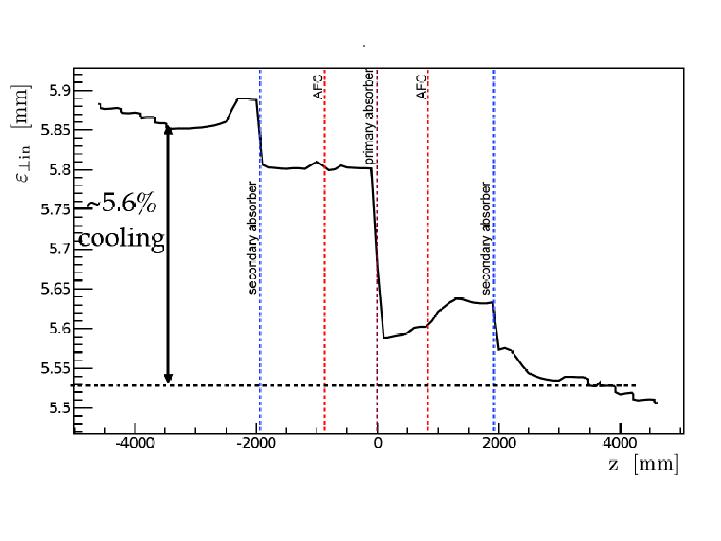}}
\label{fig-mice}
\end{center}
\vskip -1cm
\caption{Top panel: view of the MICE experiment at RAL (for more details
see \cite{mice1}). The cooling channel
is put between two magnetic spectrometers \cite{Ellis} 
and two TOF stations \cite{yordan}
to measure particle parameters. Bottom panel: 
evolution of the 4D emittance in the MICE ionization-colling demo lattice,
for a $6 \pi \cdot$ mm, 200 MeV/c muon beam.}
\end{figure}
A dedicated muon beam from ISIS (140-240 MeV/c  momentum, 
tunable between $3-10 \pi \cdot  $ mm rad  input emittance) enters
the MICE cooling section after a Pb diffuser of adjustable thickness. 
The MICE beamline has been characterized  
by the use of the TOF detectors (with $\sim 50$ ps resolution),
with data taken mainly in summer 2010 \cite{Adams}. 
As conventional emittance measurement techniques reach barely a $10 \%$ 
precision, 
the final measure of emittance will be done in  MICE   
on a particle-by-particle basis   
by measuring $x,y,x'=p_x/p_z,y'=p_y/p_z,E,t$ with the trackers
 and the TOF system. 
Foreseen performances of the MICE cooling cell are shown in the bottom panel 
of figure 2.

\subsubsection{6D cooling}
\begin{figure}[htbp] % figures (and tables) should go top or bottom of
                    % the page where they are first cited or in
                    % subsequent pages
\centering
\includegraphics[scale=0.22]{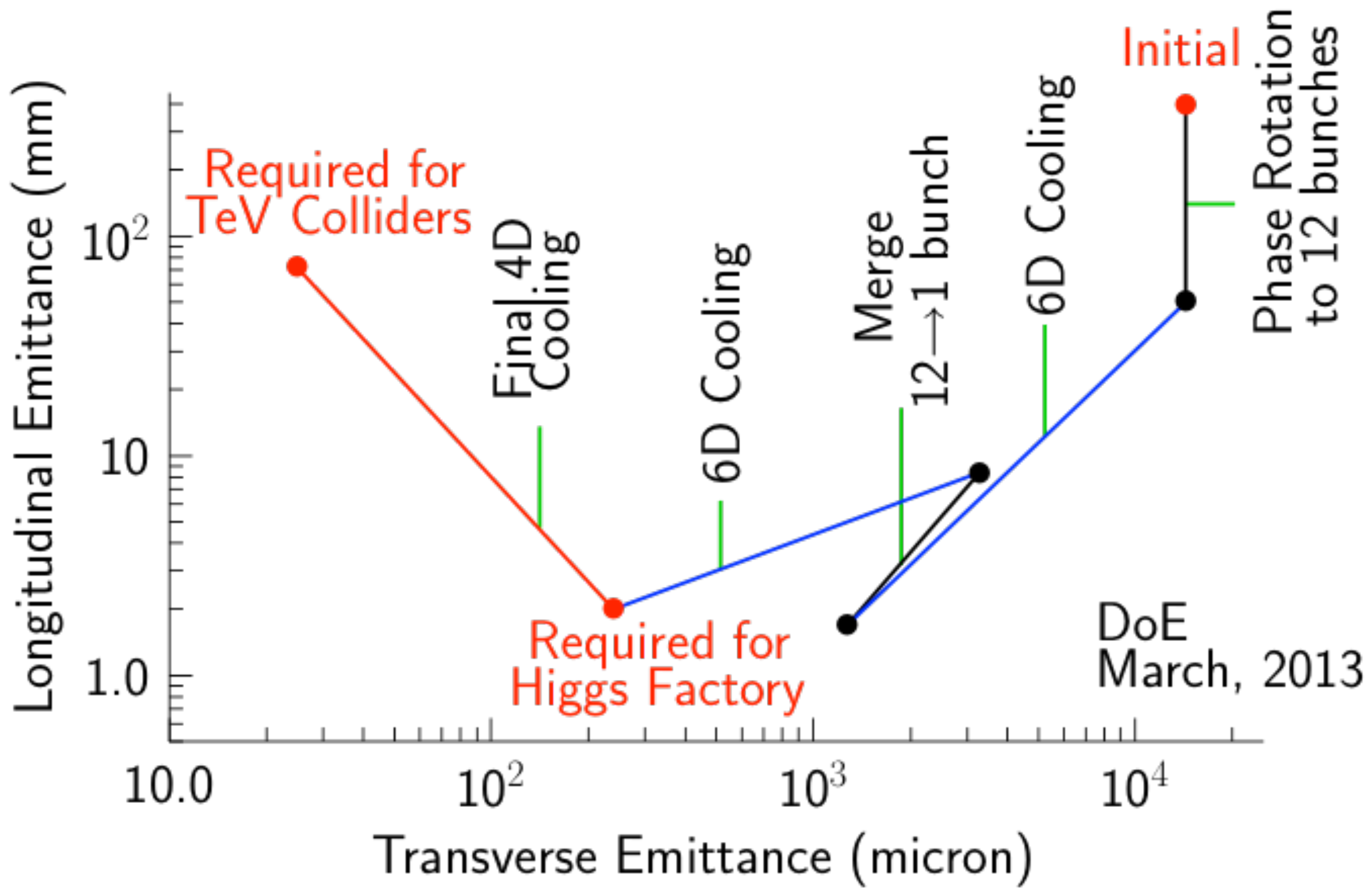}
\hspace{1cm}
\includegraphics[scale=0.27]{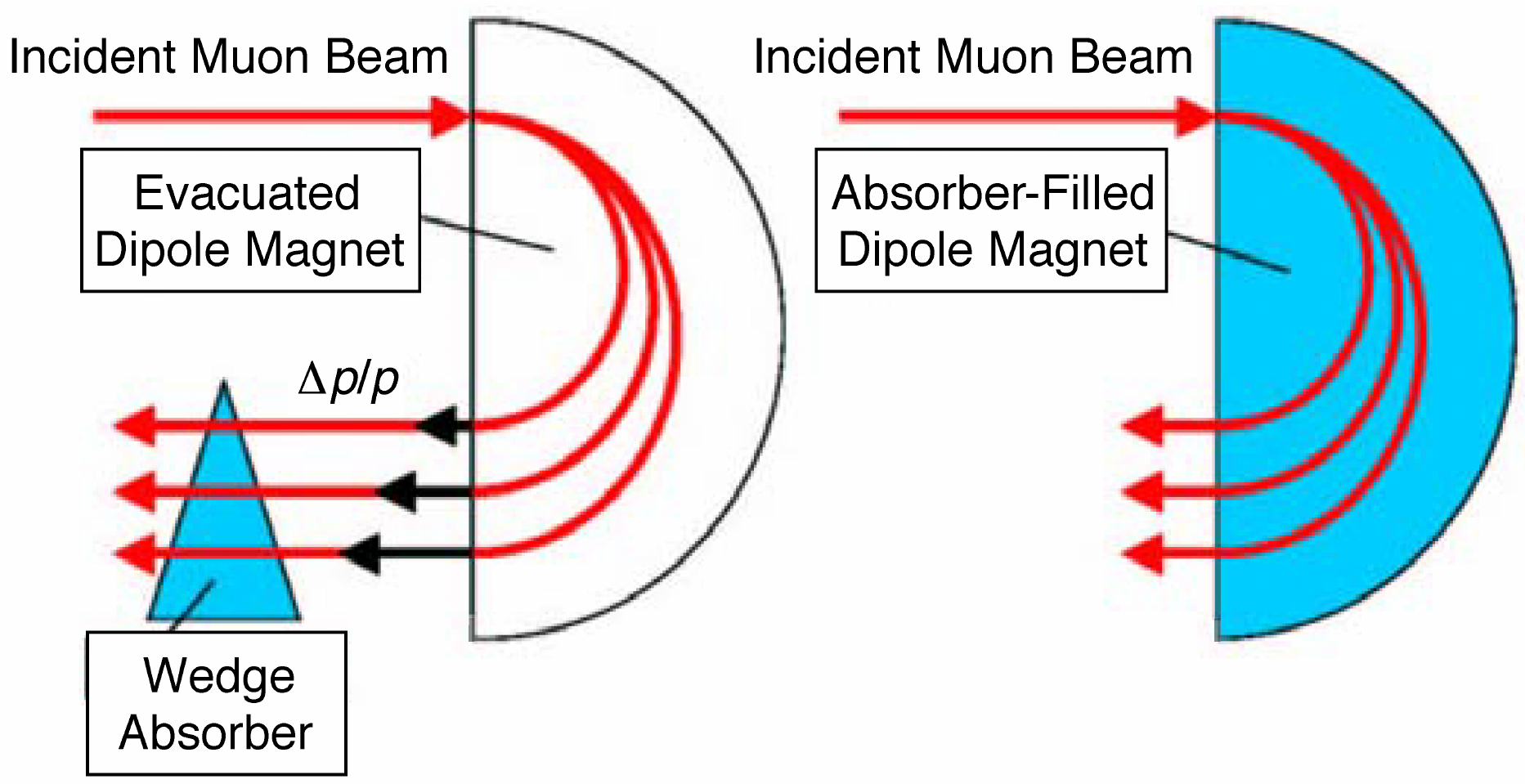}
\caption{Left panel: emittance evolution path for a $\mu^+\mu^-$ Higgs 
factory 
and a multi-TeV collider. Right panel: approaches to emittance exchange, 
to get 6D cooling [courtesy of Muons Inc.].
}
\label{fig:wedge}
\end{figure}

Both a reduction in transverse emittance
and longitudinal emittance are needed for a $\mu^+\mu^-$ Higgs factory
or a multi-TeV collider,
as shown in the left panel of figure \ref{fig:wedge} 
from reference \cite{Palmer13}. 
As a direct longitudinal cooling is not feasible,
due to the energy-loss straggling that increases the energy spread,
%%\cite{Neuffer3}, \cite{Fernow}.
the only practical solution is to transfer a fraction of the cooling 
effect from 
transverse to longitudinal phase space (via ``emittance exchange''),
as shown schematically in figure \ref{fig:wedge}. 
Dispersion is used to create an appropriate correlation between momentum
and transverse position/path length.
Clearly this is
at the expense of a reduced transverse cooling. 
Some aspects of the ``emittance exchange '' will be addressed also in
the  MICE experiment, by inserting LiH wedge absorbers.

One may envisage multi-pass cooling rings \cite{Fernow1}
 and then extract the cooled beams, with a substantial cost reduction,
instead of single-pass linear cooling channels,  as in MICE. 
These designs are based on solenoidal focussing strictly interleaved with RF 
accelerating cavities \cite{Garren}, \cite{Babelkov}, \cite{Palmer2}. 
Difficult beam dynamics must be handled and  performance limits  
or cost-effectiveness
are not completely defined. 
In a multi-turn cooling ring, the main problems will be connected
to beam injection and extraction. 

\section{Conclusions}
The recent discovery of the Standard Model Higgs at about 126 GeV has 
revived the interest for a compact muon collider: the Higgs-factory. 
As cooling factors up to $10^6$ are needed for a MC,
the optimization of the cooling channel is essential.
 A vigorous $ R \& D$  program is thus 
needed.

\end{document}